# Competitive Solvation of (Bis)(trifluoromethanesulfonyl)imide Anion by Acetonitrile and Water

Vitaly Chaban[1]

MEMPHYS — Center for Biomembrane Physics, Syddansk Universitet, Odense M., 5230, Kingdom of Denmark

**Abstract**. Competitive solvation of an ion by two or more solvents is one of the key phenomena determining the identity of our world. Solvation in polar solvents frequently originates from non-additive non-covalent interactions. Pre-parametrized potentials poorly capture these interactions, unless the force field derivation is repeated for every new system. Development cost increases drastically as new chemical species are supplied. This work represents an alternative simulation approach, PM7-MD, by coupling the latest semiempirical parametrization, PM7, with equation-of-motion propagation scheme and temperature coupling. Using a competitive solvation of (bis)(trifluoromethanesulfonyl)imide anion in acetonitrile and water, the work demonstrates efficiency and robustness of PM7-MD.

---

[1] E-mail: vvchaban@gmail.com

TOC Graphic

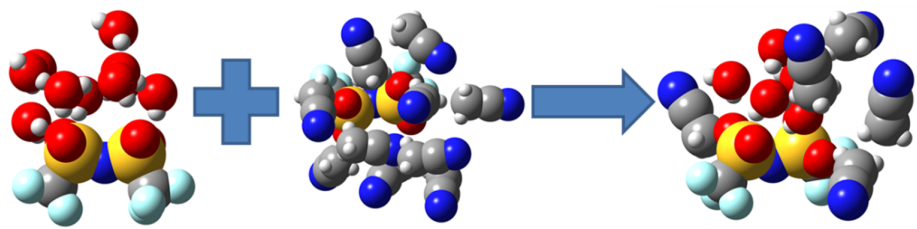

**Introduction**

Room-temperature ionic liquids (RTILs) constitute currently an emerging research field.[1-12] Applications of RTILs range from high-performance electrolyte solutions[6] to green solvents[9, 11] to separation setups,[4, 9] and poisonous gas capture.[3] RTILs are composed of bulky organic cation and inorganic or, more rarely, organic anion. An aromatic ring based cation (*N*, *N'*-dialkylimidazolium, *N*-alkylpyridinium, alkylsulphonium, thiazolium) must be asymmetric in shape, while an anion (tetrafluoroborate, hexafluorophosphate, bis(trifluoromethylsulfonyl)imide, acetate, dicyanamide) can be symmetric. Inability of the cation and anion to form an energetically efficient lattice prevents their freezing at high temperatures, such as it occurs in the case of ionic inorganic compounds (NaCl, KCl, etc). The latter form ionic melts above melting temperature, which can also be regarded as ionic liquids, but not as *room-temperature* ionic liquids.

RTILs are frequently used in combination with molecular co-solvents (water, acetonitrile, acetone, alcohols, etc),[13-15] which allow for robust modification of their physical chemical properties. An alternative example is biphasic systems for separation applications. Understanding of ion-molecular interactions in the ionic liquid-molecular liquid systems is extremely important to shape the future in this versatile field. For instance, many RTILs are notably hygrophilic. Their minimal contact with moisture may not only result in favorable changes (for certain applications), but also in the complete alteration of the desirable set of properties. Impact of the non-aqueous solvents is even less studied, especially when it comes to a systematic description.

Molecular dynamics simulations employing phenomenological potentials for non-covalent binding and harmonic potentials for covalent bonds provide invaluable insights into atomistic-resolution organization and transport properties of the RTIL containing systems. Unfortunately, phenomenological potentials are rigorously non-transferable and must, therefore, be re-parametrized for each new system. Parametrization cost increases exponentially with the number of involved chemical species. This feature makes it impossible to simulate many-component systems in practice, unless significant simplifications are introduced and certain types of interactions are excluded.

In this work, PM7[16] powered molecular dynamics (MD), PM7-MD, is applied to the investigation of (bis)(trifluoromethanesulfonyl)imide anion (TFSI⁻) in water ($H_2O$), acetonitrile (ACN), and their equimolar mixture (Figure 1). TFSI⁻ is found in lots of RTILs,[13, 15, 17] which are important in pure and applied chemistry. The anion exceeds most of other RTIL anions by size and contains a few elements which may be a coordination site (nitrogen, oxygen, sulfur) for the molecules of polar molecular co-solvent. Carbon and fluorine atoms, composing the two trifluoromethyl radicals, are hydrophobic in this configuration. However, they can contribute to the entropic component of the solvation free energy and, in this way, influence preferential solvation of TFSI⁻. Apart from introducing factual physical chemical insights, this work argues that PM7-MD presents the most cost efficient approach to address competitive solvation of ions.

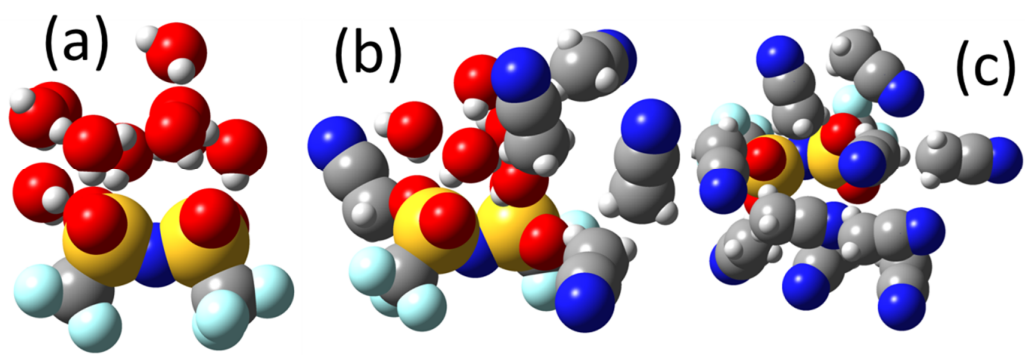

Figure 1. Optimized molecular geometries for the solvated TFSI⁻ anion (a) in pure water, system #1; (b) in equimolar water-acetonitrile mixture, system #2; (c) in pure acetonitrile, system #3.

**Methodology**

The provided analysis is based on the molecular dynamics trajectories, which were recorded using an electronic structure description of the nuclear-electronic system. That is, the wave function of all given molecular configurations was optimized repeatedly using the self-consistent field methods (SCF, convergence criterion of $10^{-7}$ Ha). The immediate forces acting on each atomic nucleus were derived from the optimized wave function at each time-step. The nuclear trajectory was propagated classically via velocity Verlet integration scheme. Both velocity and position were calculated at the same value of the time variable. The implemented methodology follows the Born-Oppenheimer approximation, which is currently widely applied in computational molecular physics.

The wave function at each nuclear time-step was computed by PM7 ("parametrized model seven") semiempirical Hamiltonian. PM7 is a quantum chemistry method,[5, 16, 18] which is based on the Hartree-Fock (HF) formalism. It is, therefore, more theoretically fundamental and robust for applications, as compared to empirical force field methods. PM7 was responsible for the electronic part of the calculation to obtain molecular orbitals, heat of formation, and its derivative with respect to molecular geometry. However, as compared to the non-parametrized HF method, PM7 makes a set of approximations and adopts certain parameters from experimental data to speed up and facilitate wave function convergence. Except serving for faster performance of SCF, empirical parameters also constitute a means to include electron correlation effects, which are principally omitted in the HF method. PM7 is currently the latest development in the family of semiempirical approaches.[16, 19] It normally offers a high accuracy of optimized geometries, thermochemistry, band gaps, and electronic spectra. PM7 favorably differs from ab initio electronic structure methods by computational cost. The performance difference comes largely from faster SCF convergence thanks to pre-parametrized integrals. PM7 routinely uses two experimentally determined constants per atom: atomic weight and heat of atomization. Electrostatic repulsion and exchange stabilization are explicitly taken into account. All applicable HF integrals are evaluated by approximate means. The set of basis functions consists of one *s* orbital, three *p* orbitals, and five *d* orbitals per each atom. Basis *d* orbitals are omitted for elements without *d* electrons. The overlap integrals in the secular equation are ignored.[16, 19]

The nuclear equations-of-motion were integrated with a 1.0 fs time-step. The simulations were performed in the constant temperature constant number of particles ensemble. The constant temperature of 300 K was maintained by the Berendsen thermostat[20] with a relaxation constant of 50 fs. In principle, the thermostat can be avoided in this system, provided that sufficiently small time-step is used for nuclear motion and the initial molecular geometry is in the local minimum state. The lengths of nuclear trajectories (Table 1) were decided on-the-fly, depending on the evolution of thermodynamics properties (Figure 2) and the system size in electrons.

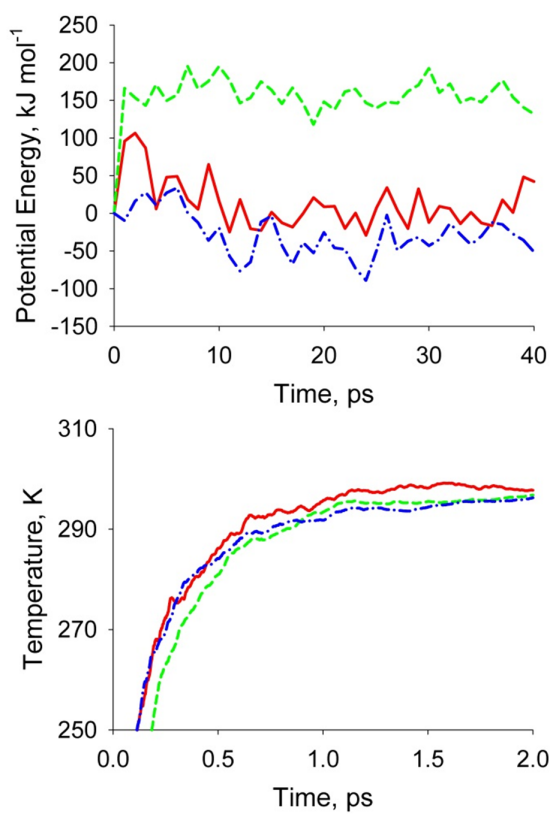

Figure 2. Evolution of temperature and potential energy during the equilibration phase of the PM7-MD simulations. The potential energy at time=0 is set 0 in all systems for clarity.

System #1 is red solid line, system #2 is green dashed line, system #3 is blue dash-dotted line. The definition of these systems is provided in Table 1.

The latest available revision of MOPAC2012 was used to obtain wave functions and forces. The Atomistic Simulation Environment (ASE)[21] set of scripts was used as a starting point to interface electronic structure stage of computations with temperature-coupled velocity Verlet trajectory propagation. The results were analyzed using the home-made programs and Gromacs package supplementary utilities.[22]

**Results and Discussion**

Figure 2 measures evolution of potential and kinetic energies vs. time. It is necessary to ensure that the system has attained thermodynamic equilibrium. Noteworthy, all three systems require a few picoseconds to achieve their proper configurations. It is irrespective of the initial geometry optimization applied to all systems. Geometry optimization alone was not able to find a proper minimum energy configuration for these electron-nuclear configurations, which are significantly large. Thermal motion perturbation (corresponding to 300 K) was necessary to overcome potential energy barriers. In the case of water-acetonitrile mixture, this occurred during the first 14 ps of dynamics. Note, that stabilization of kinetic energy occurs within the first picosecond upon the Berendsen temperature coupling. The observed changes in the potential energy at zero-temperature and 300 K are, therefore, not only entropy induced, but also enthalpy induced.

Table 1 introduces the simulated systems and the most important parameters of their setups. All three systems consist of ten co-solvent molecules and a single TFSI⁻ anion. The total number of electrons is significantly different though. Therefore, the computational cost of each simulation is different. The total sampling time was set in accordance with accumulation of sufficient statistics and approximately equivalent computational cost of all simulations.

Table 1. The list of the simulated systems. Provided is a total number of electrons per system for a straightforward comparison of computational load with alternative electronic structure studies. Note, that PM7 uses effective-core potentials for all elements except hydrogen and helium. Equilibration time was computed from evolution of thermodynamics properties in Figure 2. The three additional simulations (omitted in the table) of 5 000 fs each for system #2 were used to investigate energy conservation during the course of PM7-MD

| # | # $H_2O$ | # $CH_3CN$ | # atoms | # electrons | Sampling time, ps | Equilibration time, ps |
|---|---|---|---|---|---|---|
| 1 | 10 | 0 | 45 | 176 | 150 | 11 |
| 2 | 5 | 5 | 60 | 246 | 120 | 14 |
| 3 | 0 | 10 | 75 | 316 | 100 | 2.0 |

Figure 3 depicts radial distribution functions (RDFs) involving a few prospective solvent coordination sites of the TFSI⁻ anion in water. RDFs for carbon and fluorine sites are also plotted, although – based on the conventional chemical wisdom – they very unlikely participate as primary solvent coordination sites due to low polarity. Indeed, carbon-oxygen RDF exhibits a modest peak at 0.53 nm, which well exceeds the sum of the van der Waals radii of these atoms. Fluorine-oxygen RDF does not exhibit a distinguishable peak (Figure 3). The most articulated water – TFSI⁻ anion peak is observed in the case of oxygen (anion) – oxygen (water) distance correlation at 0.27 nm. The height of this peak is 21 units. The conjugated peak, oxygen (anion) – hydrogen (water), is located at 0.17 nm, whereas its height is 10 units. The energetically favorable separation of 0.17 nm indicates hydrogen bonding between the anion and the solvent molecule. Interestingly, the participation of imide nitrogen is mediocre, irrespective of its large partial electrostatic charge. To recapitulate, the TFSI⁻ anion is strongly coordinated in the aqueous solution. This feature is necessary to understand why the TFSI⁻ containing RTILs are water soluble and why water influences their physical chemical properties so drastically.

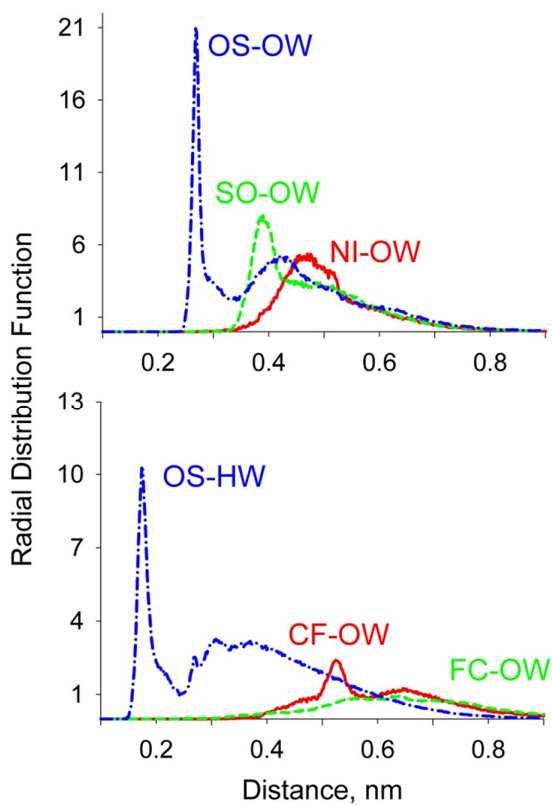

Figure 3. Radial distribution functions for the selected sites of the TFSI$^-$ anion and surrounding water molecules. Atom names in the legends are used as in common empirical force fields: OS – oxygen of TFSI$^-$; SO – sulfur of TFSI$^-$; NI – nitrogen of TFSI$^-$; CF – carbon in trifluoromethyl group of TFSI$^-$; FC – fluorine in trifluoromethyl group of TFSI$^-$; OW – oxygen of water; HW – hydrogen of water.

Figure 4 provides RDFs for TFSI$^-$ – ACN. Positively (CH$_3$) and negatively (N) charged sites of ACN molecule are investigated in relation to the most polar atoms (SO, OS, NI) of the anion. Negatively charges sites of acetonitrile molecules do not coordinate TFSI$^-$. All peaks are located at the distances, which clearly exceed the sums of the van der Waals radii of the corresponding atoms. Instead, the coordination is performed by the methyl group, CH$_3$. The TFSI$^-$ oxygen – ACN CH$_3$ peak is located at 0.30 nm with a height of 13 units. This peak is broader and of smaller height as compared to analogous peaks in water (Figure 3). Unlike in water, hydrogen atoms of acetonitrile are bound to carbon atom. That is, they cannot obtain a sufficient positive charge to initiate hydrogen bonding of the same strength as in the other cases (water, alcohols, carbonic acids, etc). In the TFSI$^-$ – ACN system, the distance between H and O equals to 0.23 nm. In turn, it equals to 0.20 nm in the optimized configuration.

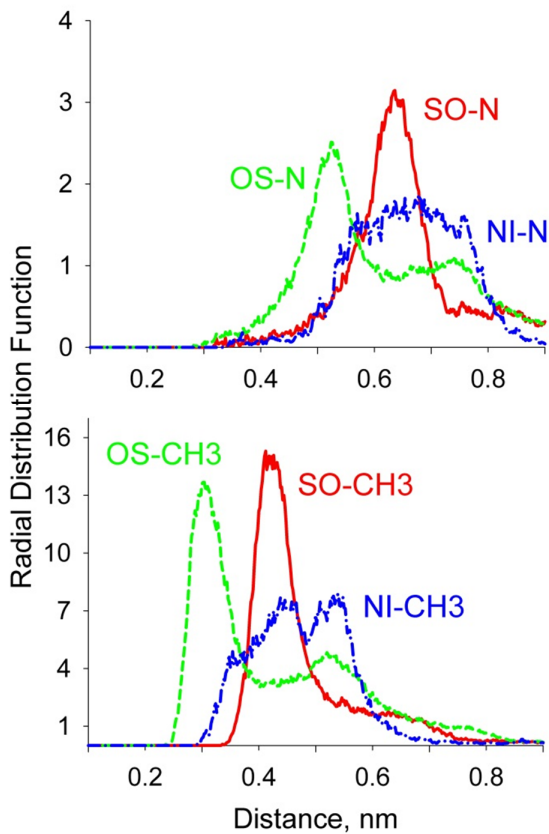

Figure 4. Radial distribution functions for the selected sites of the TFSI$^-$ anion and surrounding acetonitrile molecules. Atom names in the legends are used as in common empirical force fields: OS – oxygen of TFSI$^-$; SO – sulfur of TFSI$^-$; NI – nitrogen of TFSI$^-$; CF – carbon in trifluoromethyl group of TFSI$^-$; FC – fluorine in trifluoromethyl group of TFSI$^-$; N – nitrogen of ACN; CH$_3$ – carbon of methyl group.

Competitive solvation of the TFSI$^-$ anion by water and acetonitrile is addressed in Figure 5. Both water oxygen – TFSI$^-$ oxygen and water hydrogen – TFSI$^-$ oxygen peaks are located at smaller distances than ACN CH$_3$ – TFSI$^-$ oxygen and ACN CH$_3$ – TFSI$^-$ oxygen peaks. An ability of water molecules to approach the anion more closely plays a decisive role in the competition with ACN molecules. Compare, the methyl group of ACN is, on the average, 0.31 nm away from TFSI$^-$ oxygen. This observation is in conflict with dipole moments of the individual ACN and water molecules. Since ACN exhibits a larger dipole moment (3.9 D), one may expect its stronger attraction to the polar moiety of the anion. It is, however, not the case in the many-body system, as simulated here. Dipole moments of the molecular solvent molecules alone cannot be used to predict preferential solvation of RTIL.

Both water and acetonitrile coordinate the same site of the anion. In the context of practical applications, it is important to find such a combination of co-solvents, each of which would coordinate various anion atoms. In addition, these solvents must be miscible with one another to a reasonable extent. The dissolution of RTIL will be promoted in this way and the ion cluster size will be decreased. It is important in chemical technology during the chase for less viscous and more conductive electrolytes.[14, 15]

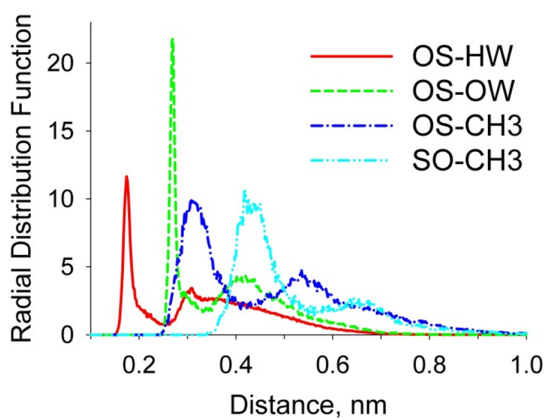

Figure 5. Competitive solvation of the TFSI⁻ anion by water and acetonitrile molecules in terms of radial distribution functions. Atom names in the legends are used as in common empirical force fields: OS – oxygen of TFSI⁻; SO – sulfur of TFSI⁻; OW – oxygen of water; HW – hydrogen of water; $CH_3$ – carbon of methyl group.

Certain comments must be provided regarding interpretation of radial distributions obtained through the PM7-MD simulations. Unlike more common studies, such as those employing classical molecular dynamics schemes and classical Monte Carlo schemes, PM7-MD does not use an approximation of periodic boundary conditions. This limitation comes from the implementation[16] of PM7 as an electronic structure method, which uses a specifically optimized set of Slater-type orbitals. The usefulness of periodic boundary conditions (PBCs) is not perfectly clear for relatively small systems, since PBCs cannot compensate for the omitted long-range ordering. In the meantime, PBCs introduce certain perturbations as a trade off for eliminating the surface effect. As long as system volume is not defined by the size of the periodic MD box, the traditional way of RDF normalization (through division per average number density in the system) cannot be used. In this work, an alternative solution was implemented. The computed RDFs were integrated over the full valid function range. This integral value should, by definition of RDF, be equal to the total number of atoms of given type divided by their number density. The total number of atoms is a constant known value. The resulting RDFs can, in such a way, be uniformly scaled to provide a correct normalization.

The heights of RDF peaks computed from PM7-MD will systematically exceed those from the condensed phase investigations. This is because a large fraction of atoms is located at the interface. These atoms do not have neighbors in all directions, meaning that the attraction is not sufficiently competitive. This does not apply to the anion itself, since it is surrounded by the coordinating solvent molecules. However, this limitation does not influence any of the meaningful conclusions, such as peak positions, preferential solvation, coordination numbers, etc. Note, that the primary goal of PM7-MD simulations is to address short-range structure. In turn, empirical molecular dynamics is a method of choice for long-range structure.[23-25]

Energy conservation in molecular dynamics simulations represents a central issue. Even though coupling to an external thermal bath is able to mask energy conservation problems, they may manifest themselves in the biased trajectories and, therefore, perturbed structure properties. Such issues are often well hidden and cannot be identified directly. Note, that unlike empirical simulations, PM7-MD does not use any constraints (such as constraints of covalent bonds involving hydrogen atoms). Therefore, all movements in the MD system are natural and the optimal integration time-step must be set in accordance with the fastest bond oscillation. Figure 6 records total energy leakage in system # 2 (Table 1) versus time. These simulations have been conducted without a thermostat (constant energy ensemble). The time-steps of 0.5 fs and 1.0 fs provide reasonable energy conservation. The changes during 5 ps do not exceed 0.2 % in both cases. On the contrary, the time-step of 2.0 fs is unsatisfactory, since the system has lost 2.5% of its initial energy during a short time period. The usage of this option would provide unphysical results. Recall, all productive simulations reported here are based on the 1.0 fs time-step.

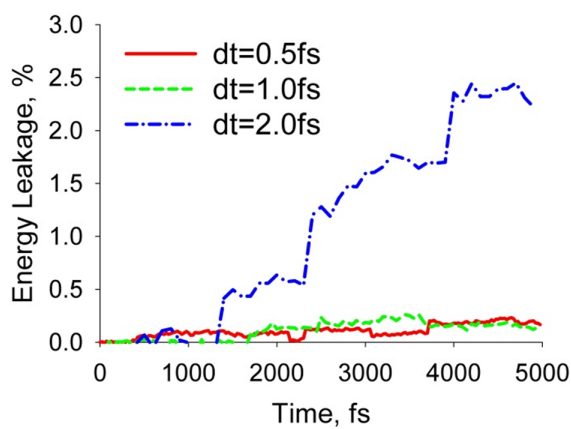

Figure 6. Total energy conservation during PM7-MD simulations using the integration time-steps of 0.5 fs (red solid line), 1.0 fs (green dashed line), and 2.0 fs (blue dash-dotted line). Whereas the time-steps of 0.5 fs and 1.0 fs provide a reasonable level of total energy conservation, the energy promptly leaks out from the system in the case of 2.0 fs time-step. The latter must be avoided in simulations, even though temperature coupling scheme may mask this problem at certain circumstances and to certain extent. Additional simulations of system #2 were used to record energy conservation at various time-steps.

**Conclusions**

PM7 powered MD simulations were, for the first time, applied to the investigation of the competitive solvation of (bis)(trifluoromethanesulfonyl)imide anion by acetonitrile and water. It was shown that water exhibits somewhat stronger affinity to the TFSI$^-$ anion due to stronger hydrogen bonding. This is despite its smaller dipole moment (1.85 D), while ACN molecule has a dipole moment of 3.9 D. Both solvent molecules coordinate four oxygen atoms of TFSI$^-$. That is, the mechanism of solvation in ACN and water is equivalent. The reported data and considerations are important to foster further progress in the emerging field of ionic liquids.

The major advantage of PM7-MD is a universal parametrization of bonded and non-bonded interactions, which depends only on chemical elements (as opposed to molecules or fragments) in the system. Consequently, the transferability of this method is higher than that of other parametrized methods. Coupled with a beneficial computational cost, it allows to sample hundreds of picoseconds for the systems containing hundreds to thousands electrons. PM7-MD, therefore, fills a niche between first-principles molecular dynamics and more phenomenological models based on empirical potentials.

**Acknowledgments**

I thank Dr. James J.P. Stewart (President at Stewart Computational Chemistry, Colorado Springs, United States) for extensive discussions, wise comments and, most importantly, inspiration. I thank University of Rochester, New York, United States and personally Prof. Oleg V. Prezhdo and Dr. Eric Lobenstine for providing me a courtesy library access outside working hours. MEMPHYS is the Danish National Center of Excellence for Biomembrane Physics. The Center is supported by the Danish National Research Foundation.

I thank Nadezhda Andreeva (St. Petersburg, Russian Federation) for she knows what.

**Author Information**

E-mail address for correspondence: vvchaban@gmail.com; chaban@sdu.dk.

Tel.: +1 (413) 642-1688.